\RequirePackage{fix-cm}

\documentclass[twocolumn]{svjour3}          
\usepackage{amsmath}
\usepackage{multirow}
\usepackage[colorlinks,linkcolor=red]{hyperref}
\usepackage{amssymb}
\usepackage{tabularx}
\usepackage{array}
\usepackage{graphicx} 
\usepackage{float} 
\usepackage{subfigure} 
\usepackage{color}
\usepackage{booktabs}
\smartqed  

\usepackage[hyphenbreaks]{breakurl}
\usepackage{graphicx}
\journalname{CGI2018} 
\begin{document}

\title{TransMRSR: Transformer-based Self-Distilled Generative Prior for Brain MRI Super-Resolution}
\author{Shan Huang$^{*}$ \and Xiaohong Liu$^{*}$ \and Tao Tan \and Menghan  Hu \and Xiaoer Wei$^{\dag}$ \and Tingli Chen$^{\dag}$ \and Bin Sheng$^{\dag}$}
\institute{
$^{*}$Equal contribution \\
$^{\dag}$Corresponding authors. \\
Shan Huang, Xiaohong Liu  \at Department of Computer Science and Engineering, Shanghai Jiao Tong University, Shanghai 200240, China. \and  Tao Tan \at Faculty of Applied Sciences, Macao Polytechnic University, Macao 999078, China. \and Menghan Hu \at Shanghai Key Laboratory of Multidimensional Information Processing, East China Normal University, Shanghai 200241, China \and Xiaoer Wei \at Institute of Diagnostic and Interventional Radiology, Shanghai Jiao Tong University Affiliated Sixth People's Hospital, Shanghai 200233, China. \\
weixiaoer\_2003@163.com
\and Tingli Chen \at Department of Ophthalmology, Huadong Sanatorium, Wuxi 214065, China. \\
chentingli1028@163.com
\and Bin Sheng \at Department of Computer Science and Engineering, Shanghai Jiao Tong University, Shanghai 200240, China.\\
shengbin@cs.sjtu.edu.cn
}
\date{}
\maketitle

\begin{abstract}
Magnetic resonance images 
(MRI) acquired with low through-plane resolution compromise time and cost. The poor resolution in one orientation is insufficient to meet the requirement of high resolution for early diagnosis of brain disease and morphometric study. The common Single image super-resolution (SISR) solutions face two main challenges: (1) local detailed and global anatomical structural information combination; and (2) large-scale restoration when applied for reconstructing thick-slice MRI into high-resolution (HR) iso-tropic data. To address these problems, 
we propose a novel two-stage network for brain MRI SR named TransMRSR based on the convolutional blocks to extract local information and transformer blocks to capture long-range dependencies. TransMRSR consists of three modules: the shallow local feature extraction, the deep non-local feature capture, and the HR image reconstruction. We perform a generative task to encapsulate diverse priors into a generative network (GAN), which is the decoder sub-module of the deep non-local feature capture part, in the first stage. The pre-trained GAN is used for the second stage of SR task. We further eliminate the potential latent space shift caused by the two-stage training strategy through the self-distilled truncation trick. The extensive experiments show that our method achieves superior performance to other SSIR methods on both public and private datasets. Code is released
at \href{https://github.com/goddesshs/TransMRSR.git}{https://github.com/goddesshs/Trans
MRSR.git}.


\end{abstract}

\section{Introduction}
\label{sec:1}
\begin{figure}[t] 
\centering 
\includegraphics[width=1\columnwidth]{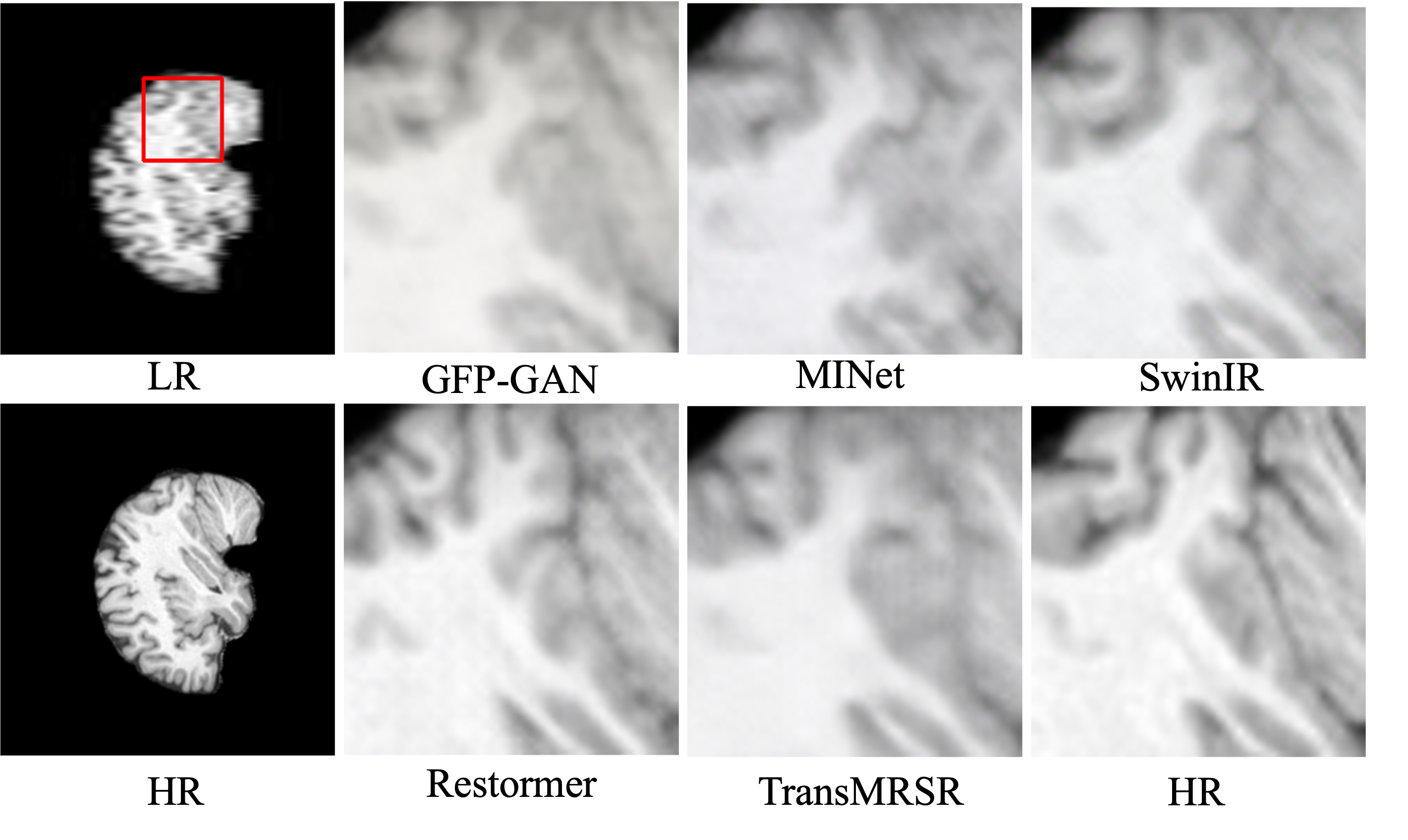} 
\caption{Visual comparison between our method and state-of-the-art SR methods on a through-plane image. The anatomical details in the reconstructed MRI image by our TransMRSR are found to be closest to those in the HR image.}  
\label{Fig.1} 
\end{figure}

Brain magnetic resonance images (MRI) are important for the early diagnosis and early treatment of brain diseases with clear anatomical structures and high contrast. Meanwhile, morphometric analysis such as accurate estimation of gray and white matter volume based on 3D brain MRI is a key technique in neuroscience to study human brain development, aging, plasticity, and disease. However, acquiring high-resolution MRI with adequate signal-to-noise (SNR) is challenging due to the prolonged acquisition procedure and patient breath-hold \cite{xia2021super}. It is common in the clinical setting to obtain anisotropic 2D MRI with high in-plane resolution($\le 1mm$), but the low through-plane resolution
($4 \sim 7mm$) in the tradeoff between quality and cost, as illustrated in Fig. \ref{Fig.1}.  Super-resolution technique is a promising post-processing tool to improve image quality without changing the MRI hardware. 

Simple use of affine transformation, \textit{e.g.}, interpolation methods, to restore isotropic voxels has serious blurriness and artifacts. The loss of most high-frequency information in the through-plane degrades accurate 3D analysis \cite{zhao2020smore}. Recently, the deep learning-based 
SR 
me-thods have shown great potential in turning a huge 
amount of routine diagnostic brain MRI into useful data for neurometric research \cite{liu2021recycling}. 2D Single image super-resolution methods work slice-by-slice to restore HR volumetric data based on 2D SR networks \cite{peng2020saint,zhang2021mr}, while 3D SISR networks are also applied to extract connections between slices \cite{chen2018brain,du2020brain,wang2020enhanced}. 
Other networks try to improve the quality of the images generated using GANs \cite{wang2020enhanced,chen2018efficient}. Although these studies have some effect, we still face two main challenges. Firstly, most existing methods harness deep convolution layers for feature extraction. The convolution kernel with the limited receptive field is inadequate to capture long-range dependencies required by complicated anatomical structure reconstruction \cite{li2022transformer}. Secondly, most networks perform well on $\times2$ or $\times4$ upsampling, yet suffer severe performance degradation on $\times8$ upsampling, which cannot satisfy the needs of real applications.

In order to address these shortcomings, we propose a transformer based 
MRI SR network named TransMRSR, an efficient approach to reconstruct 3D MRI from multiple LR through-plane slices. Our training process follows a two-stage paradigm. Firstly, we pre-train a generative StyleSwin \cite{zhang2022styleswin} by utilizing many existing HR images to provide a rich generation prior. Then, we train the whole network in the downstream super-resolution task.
Specifically, we use the StyleSwin as the decoder part of the deep feature capture module and adapt the priors to brain MRI SR tasks. We align the latent space of the decoder in the first generative task and the second SR task by self-distilled "truncation trick". Besides, we introduce residual learning between the shallow feature and deep feature modules to better recover high-frequencies better. 

In summary, Our contributions can be listed as follows:
\begin{itemize}
    \item We propose a novel architecture named TransMRSR utilizing convolution module to extract local features and transformer to capture long-range dependencies. Our TransMRSR consists of three submodules: the shallow feature extraction based on the convolution layer, the deep feature capture based on the UNet, and the HR image reconstruction. 
    \item We pre-train a StyleSwin using the existing public multi-modal HR brain MRI dataset and further use the brain prior to the MRI super-resolution task. We use the encoder to extract multi-scale features. The highest-level semantic features are converted into latent vectors to condition the style of the image. The self-distilled "truncation trick" is proposed to eliminate latent space offsets between generative and super-score tasks. In order to enforce restrictions on what the decoder generates, we use features from each layer of the encoder to adjust the output of the decoder image at multiple levels. 
    \item We perform extensive experiments on synthetic and clinical datasets. Our TransMRSR 
    outperforms 
    other SISR methods on quantitative metrics and visual quality.
\end{itemize}
\section{Related Work}
\label{sec:2}
\subsection{MR Image Super-resolution}
With the introduction of SRCNN \cite{dong2015image}, deep neural networks 
outperform traditional interpolation-based and reconstruction-based methods on various SR tasks \cite{liu2018robust,liu2020end,wang2021single,shi2021learning,liu2021exploit,chu2022nafssr}. Many studies further change the convolution neural network (CNN) architecture to improve performance \cite{kim2016accurate,shi2016real}. Using the mean square error as the objective function to train the network can achieve a high peak SNR, but the restored images usually lack high-frequency detail information and visual quality. To solve this problem, Ledig \textit{et al.} \cite{ledig2017photo} introduce the adversarial loss based on the generative adversarial network (GAN). A common issue of the convolution network is that the limited receptive field of the convolution kernel cannot adequately capture long-range dependencies. Liang \textit{et al} \cite{zhang2022styleswin} combine the local features extracted by the convolution layer and non-local features extracted by the Swim Transformer block, and obtain better performance with fewer parameters.   

\begin{figure*}[t] 
\centering 
\includegraphics[width=1.9\columnwidth]{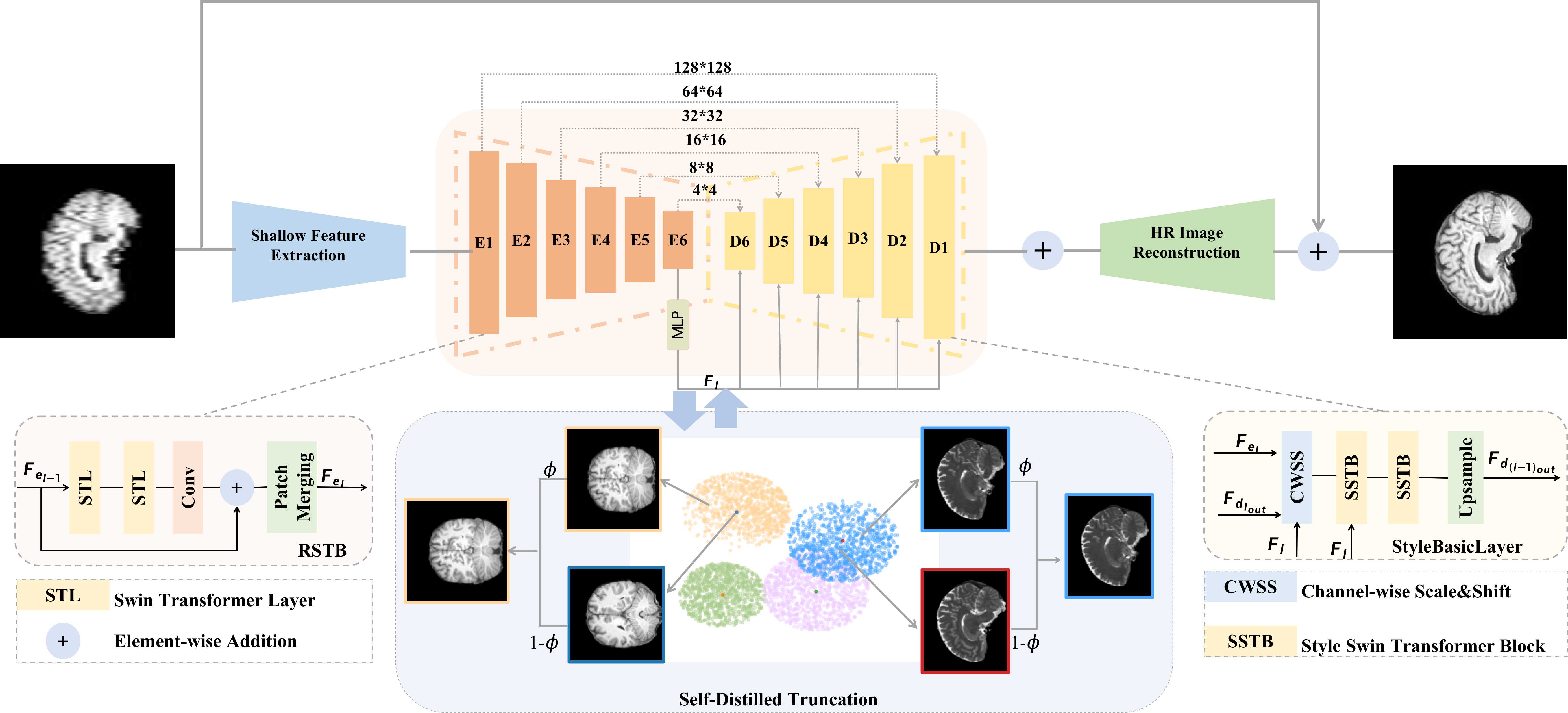} 
\caption{The overall architecture of the proposed TransMRSR. Each encoder layer is a residual swim transformer block (RSTB), which consists of two swim transformer layers (STL). Decoder is a StyleSwim network containing several Stylelayer. SSTB denotes the StyleSwimTranfromer block conditioned with latent vectors from the output of the decoder. The latent codes extracted by the encoder are truncated with the nearest centroid which is computed after the generative task.} 
\label{Fig.2} 
\end{figure*}

The success of deep learning-based natural SISR methods promotes the application of deep learning technologies in MR image SR tasks \cite{zhang2021mr,chen2018brain,du2020brain,li2022transformer,feng2021multi,lyu2020mri,zhang2019mri}. For instance, Du \textit{et al.} \cite{du2020brain} use long and short skip connections for brain MRI reconstruction. Lyu \textit{et al.} \cite{lyu2020mri} feed five generic super-resolution models into GAN based on ensemble learning to obtain the final super-resolution image. Zhang \textit{et al.} \cite{zhang2021mr} propose squeeze and excitation reasoning attention networks, which recalibrate feature responses with adaptive attention vectors learned by primitive relationship reasoning attention.  Considering the information from continuous structure, the 3D convolution networks are also used for the MR image SR tasks \cite{chen2018brain,du2020brain,zhang2019mri}. Utilizing the complementary information between MRI multi-contrast images is a promising way to yield SR images with higher information \cite{feng2021multi,li2022transformer}. Li et.al \cite{li2022transformer} develop innovative Transformer empowered multi-scale contextual matching and aggregation techniques. A common issue for these methods is finding the connection between the reference image and the low image, which is difficult or time-consuming.

\subsection{Generative Prior}
GAN, composed of a generator and a discriminator, restores high-frequency details and produces perceptually satisfying SR images \cite{chartier2018multi,jin2022deep}. Recent studies have shown that GANs effectively encode rich semantic information in intermediate features \cite{bau2020semantic} and latent codes \cite{goetschalckx2019ganalyze}. GAN inversion aims to invert a given image back into the latent code of a pre-trained GAN model \cite{xia2022gan}. PLUSE \cite{menon2020pulse} iteratively optimize the latent code of StyleGAN for each input. mGANprior \cite{gu2020image} employs multiple latent codes and integrates them with adaptive channel importance to recover the input image. The optimization-based methods typically require large memory and long time to find the closest latent code. Zhu \textit{et al.} \cite{zhu2020domain} train a separate encoder to obtain latent code $z$ as the initialization for optimization. However, the low-dimension latent codes are insufficient to keep faithful spatial information. GLEAN \cite{chan2021glean} conditions the pre-trained styleGAN with latent code and multi-resolution convolutional features. GFPGAN \cite{wang2021towards} extracts latent codes and multi-scale features based on the U-Net structure and then feed them into pre-trained StyleGAN2.
Korkmaz \textit{et al.} \cite{korkmaz2022unsupervised} learns a high-quality MRI prior in an unsupervised generative modeling task and optimize zero-shot reconstruction objective.
\section{Method}
\subsection{Overall Architecture}
Consider an MRI volume $V(x,y,z) \in \mathbb{R}^{x \times y \times z}$,
we refer to the $x$ axis as the sagittal axis, the $y$ axis as the 
coronal axis, and the $z$ axis as the axial axis. $x$ is equal to $y$ much greater than $z$ for an anisotropic volume taken along z axis. 
The aim of this work is to restore high-resolution (HR) image $V(x,y,z)\in \mathbb{R}^{x\times y \times r \cdot z}$ from low-resolution
(LR) image, where $r=x/z \gg 1$. 

The overall architecture of TransMRSR is depicted in Fig. \ref{Fig.2}. Our TransMRSR consists of three parts: the shallow feature extraction based on the convolution layer, the deep feature capture based on the UNet, and the HR image reconstruction. The shallow feature extraction and HR image reconstruction modules are composed of several residual convolutional blocks. We \textbf{design} the deep features extraction module based on the encoder-decoder architecture. We first perform a generative task based on the StyleSwin to gather diverse priors. Next, we train the whole network for the downstream SR task. To be more precise, we utilize it as the decoder of the deep feature capture module and fine-tune the parameters to adapt to the SR task. Given an interpolated low-resolution image as input, 

TransMRSR performs the restoration process as follows: 
\begin{itemize}
    \item TransMRSR first applies several residual blocks to obtain shallow feature embeddings $F_{s}$.
    \item Then, these low-level features pass through a symmetric encoder-decoder and are transformed into deep features. The encoder receives the low-level feature and exponentially reduces the size of the feature maps to $4 \times 4$. After that, the high-level hint $F_{e}$ is mapped to intermediate latent vectors through one linear layer. We further eliminate the potential latent space shift caused by the two-stage training strategy through the self-distilled truncation trick before passing them to the decoder.
    Specifically, the latent vectors are interpolated with the nearest centroid which is computed once after the generative task. Starting from a learned constant input, the pre-trained decoder adjusts the style of the image at each convolution layer based on the latent code \cite{karras2019style}. To further improve reality and fidelity, multi-resolution encoder features $F_{e}$  are used to modulate the decoder features $F_{d}$ through the Channel-Wise Scale\&Shift layer (CWSS). The decoder successively recovers the high-resolution representations. 
    \item Next, HR reconstruction layers aggregate the low-level local features extracted by convolution operation and long-range dependencies captured by Transformer based module to generate a residual image. Both local and global features are well fused to preserve structural and textual details in the restored images.
    \item Finally, the degraded image and the residual image are added to generate the final output. 
\end{itemize} 

\subsection{Encoder}
After the input LR image passes through convolution layers to get shallow feature maps $F_{s}$, \textit{i.e.},  $F_{e_{0}}$, we apply several hierarchical encoder layers to reduce the size of feature maps as:
\begin{equation}
    F_{e_{l}} = E_{l}(F_{e_{l-1}})
\end{equation}
residual Swin Transformer Block (RSTB) followed by a $3\times3$ 
convolution layer for feature enhancement. We adopt shortcut connections between input and out to stabilize feature extraction. The feature map size is reduced by the patch merging layer, which also diversifies windows. Specifically, given the input feature map $F_{e_{l-1}} \in \mathbb{R}^{H \times W \times C_{1}}$ of layer $l-1$, the Encoder operation is as follows:
\begin{align}
    \hat{F_{e_{l}}} &= Conv(RSTB(F_{e_{l-1}})) + F_{e_{l-1}} \\
    {F_{e_{l}}} &= PM(\hat{F_{e_{l}}})
\end{align}
where PM denotes the patch merging layer. A PM concatenates the features of each group of $2 \times 2$ neighboring patches to reduce the size of the feature map from $H \times W $ to $\frac{H}{2}  \times \frac{W}{2}$ and applies a linear layer increase the number of channels. Finally, we let the output of the last layer pass through a fully-connected layer to get the latent vectors. For fewer artifacts, the vector input to each decoder block is different from the same vectors used by StyleSwin:
\begin{equation}
    F_{l} = MLP(F_{e_{N}})
\end{equation}
The latent features $F_{l}$ guide the pre-trained decoder to generate an HR image with high-level information. To make full advantage of the multi-scale features produced by the encoder, we further merge the output of every encoder layer into the corresponding decoder layer.
\label{sec:3}

\begin{figure}[t] 
\centering 
\includegraphics[width=0.8\columnwidth]{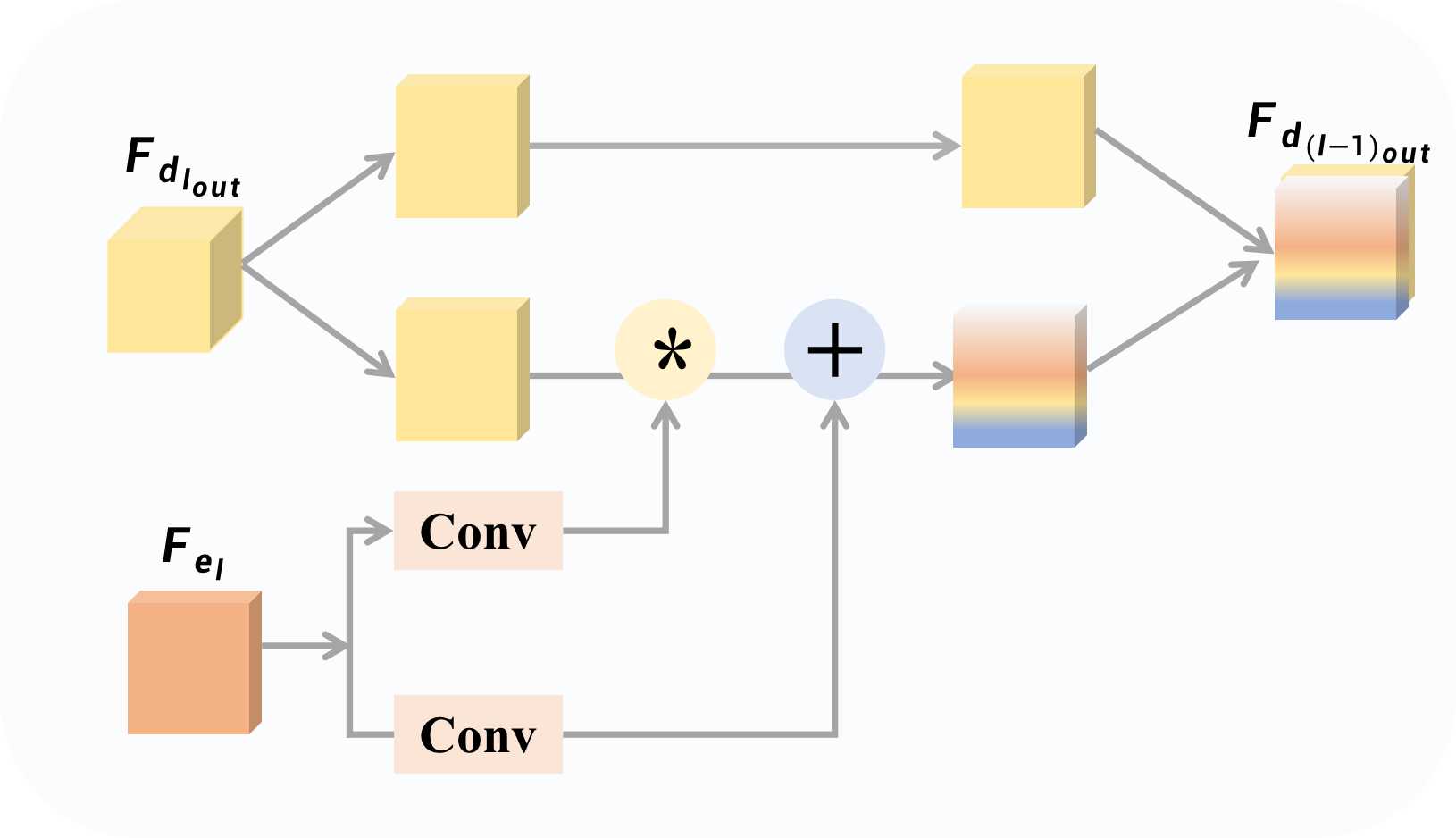} 
\caption{Channel-Wise Scale\&Shift layer} 
\label{Fig.3} 
\end{figure}

\subsection{Generative Prior Decoder}
The brain has a relatively fixed structure like the face. Inspired by \cite{chan2021glean,wang2021towards}, we pre-train a brain GAN using large-scale HR brain images to capture a distribution over the brain. Instead of exhaustively searching in a generative model's latent space to generate realistic outputs \cite{menon2020pulse}, we feed the semantic features extracted by the encoder into the decoder with self-distilled truncation and propose CWSS modules to adjust the GAN feature with the multi-resolution output of the encoder at each resolution level. We align the latent space between the generative task and the SR task by avoiding the latent code residing in the distant and sparse margins of the distribution. More specifically, we cluster $M$ randomly sampled latent codes $\left \{ F_{l_{i}} \right \} _{i=1}^{M}$ into N clusters, obtaining N cluster centers $\left \{ F_{l_{c_{j}}} \right \} _{j=1}^{N}$ \cite{mokady2022self}. This operation is performed only once. During the training and inference phase, the latent code extracted by the encoder is interpolated with the nearest centroid $F_{l_{c}}$ as:
\begin{equation}
    F_{l_{t}} = \phi F_{l} + (1-\phi) F_{l_{c}}
\end{equation}
where $\phi$ controls the truncation level. The truncated latent code pass through a stride of $1$, $3 \times 3$ Conv2D to get two parameters $\alpha$ and $\beta$. $\alpha$ is used as attention maps to scale the output of each layer of the decoder and $\beta$ is utilized to promote the input to the next layer as:
\begin{align}
    \alpha, \beta &= Conv(F_{e_{l}}) \\
    F_{d_{(l-1)_{in}}} &= \alpha \odot F_{d_{l_{out}}} + \beta
\end{align}
where $F_{d_{(l-1)_{in}}}$, $F_{d_{l_{out}}}$ denotes the input of the
$(l-1)th$ layer and the output of the $l$-1 layer of the encoder, $\odot$
denotes channel-wise multiplication. For a balance of realness and fidelity, we perform scale and shift on part of the GAN features and leave the left features unchanged, as shown in Fig. \ref{Fig.3}:
\begin{align}
    F_{d_{(l-1)_{in}}} &= CWSS(F_{d_{l_{out}}}|\alpha, \beta) \\
                   &= Concat[F_{d_{l_{out}}}^{0}, \alpha \odot F_{d_{l_{out}}}^{1} + \beta]
\end{align}
where $F_{d_{l_{out}}}^{0}$, $F_{d_{l_{out}}}^{1}$ are the split features from $F_{d_{l_{out}}}$ in channel dimension, and $Concat \left [\cdot ,\cdot  \right ] $ denotes the concatenation operation \cite{wang2021towards}.
Then, $F_{d_{(l-1)_{in}}}$ passes
through styleBasic-layer of depth $2$. The block split attention heads into two groups, one for window-based multi-head self-attention and the other for shifted-window-based self-attention. Finally, we adopt bilinear interpolation for upsampling. Details can be found in \cite{zhang2022styleswin}.

\subsection{Skip Connection}
We design skip connection following \cite{liang2021swinir}. The shallow features and deep features are fused before passing thro-ugh the HR image reconstruction part as:
\begin{equation}
    I_{Res} = H_{Rec}(F_{s} + F_{d_{1}})
\end{equation}
where $H_{Rec}(\cdot)$ is the function of HR image reconstruction module. The shallow feature layers are responsible for extracting low-frequency and local information such as edges and textures, while the UNet network specializes in capturing high-frequency information and long-range dependencies. The TransMRSR takes it a step further by utilizing a long skip connection that allows the low-frequency information to be directly transmitted to the reconstruction module. This strategy enables the deep feature extraction module to concentrate on high-frequency information and also stabilizes the training process.
We also apply residual learning to reconstruct the residual between the LR and HR image instead of the HQ image as:
\begin{equation}
    I_{SR} = I_{LR} + I_{Res}
\end{equation}
where $I_{SR}$  denotes the reconstructed MR image.

\begin{table*}[htbp]
\renewcommand\arraystretch{1.5}

\centering
\caption{Quantitative results on the IXI-Test and SixP dataset with different enlargement scales, in terms of PSNR and SSIM. The best and second-best are marked in red and blue, respectively.} 
\label{table:results} 
\begin{tabular}{r|cc|cc|cc|cc}
\hline
Dataset   & \multicolumn{4}{c|}{IXI-Test}   & \multicolumn{4}{c}{SixP}       \\
\hline
Scale     & \multicolumn{2}{c|}{$\times$4}&  \multicolumn{2}{c|}{$\times$8} &  \multicolumn{2}{c|}{$\times$4}  &  \multicolumn{2}{c}{$\times$8}  \\
\hline
Metrics   & PSNR  & SSIM   & PSNR  & SSIM   & PSNR  & SSIM    & PSNR  & SSIM   \\
\hline
\hline
EDSR      & 34.14 & 0.9574 & 29.67 & 0.9209 & 33.52      & 0.9437        & 27.36      & 0.8561       \\
GFP-GAN   & 36.37     &  0.9753      & 29.50      &  0.9108      &  35.29     &    0.9683     & 29.15      &  0.9002      \\
MINet    &   38.04    &   0.9826     &   30.10    &  0.8923      &  36.31     &   0.9741      &   29.09    & 0.8758     \\
SwinIR    & \textcolor{blue}{37.84} & \textcolor{blue}{0.9837} & \textcolor{blue}{30.47} & 0.9371 & \textcolor{blue}{36.42} & \textcolor{blue}{0.9764 } & \textcolor{blue}{29.76} & \textcolor{blue}{0.9238} \\
Restormer & 37.52 & 0.9834 & 30.35 & \textcolor{blue}{0.9436} & 35.80     &  0.9745       & 29.31 & 0.9265 \\
\hline
\textbf{TransMRSR}   & \color{red}38.56 & \color{red}0.9862 & \color{red}30.85 & \color{red}0.9451 & \color{red}36.98 & \color{red}0.9793 & \color{red}30.32 & \color{red}0.9353 \\
\hline
\end{tabular}
\end{table*}

\begin{figure*}
\begin{center}
    \includegraphics[width=1.9\columnwidth]{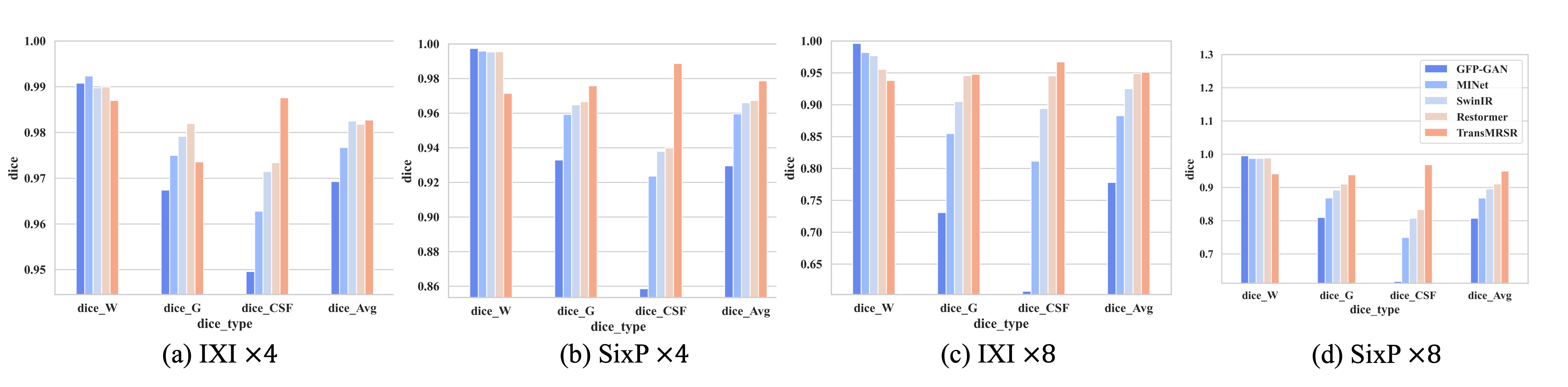}
\end{center}
\caption{Dice coefficient scores of different SR methods on two different test datasets with $\times 4$ and  $\times 8$ enlargement.}
\label{fig:dice}
\end{figure*}

\subsection{Loss Function}
\subsubsection{Reconstruction Loss}
we use L1-loss to restore the general details of SR images:
\begin{equation}
    \pounds _{rec} = \left \| I_{SR} - I_{HR} \right \|_{1}
\end{equation}
where $I_{HR}$ denotes the ground-truth.

\subsubsection{Content Loss}
To prevent over-smoothed SR results brought by L1-norm, we compare the difference between SR and HR 
in high-level feature space to improve visual quality. Particularly, feature maps from the 2nd, 4th, 7th, and 10th from the pre-trained VGG19 model are selected to compute the content loss following \cite{johnson2016perceptual}. The content loss can be viewed as the Mean-square-error (MSE) between two feature maps:
\begin{equation}
    \pounds_{cont} = \sum_{l=1}^{4}\mathbb{E}_{(I_{SR}, I_{HR})}\left \| \phi_{l} \left ( I_{SR} \right )- \phi_{l} \left ( I_{HR} \right ) \right \|_{2} ^{2}
\end{equation}
where $\phi$ denotes output the $l$-th layer of VGG19.

\subsubsection{Style Loss}
In addition to L1-loss and content loss, we also include style loss proposed by Gatys \textit{et al.} \cite{gatys2016image}. Matching extracted features statistically by computing MSE between the Gram matrices of the reconstructed image retains the texture information:
\begin{equation}
    \pounds_{style} = \mathbb{E}_{(I_{SR}, I_{HR})}\left \| G(\phi_{l} \left ( I_{SR} \right ))- G(\phi_{l} \left ( I_{HR} \right )) \right \|_{2} ^{2}
\end{equation}
where $G(F)=FF^{T} \in  \mathbb{R}^{nc\times ml}$ computes the outer product of feature map matrix $F$ and its transpose matrix $F^{T}$. $nc$ represents the number of feature maps and $ml$ is the product of the height and width of the feature maps. The Gram matrix measures the correlation between feature maps and can be used to represent the style of an image. Using this loss function promotes feature matching considering the structural coherence of brain images.

In the end, the total objective of the TransMRSR model is:
\begin{equation}
    \pounds_{total} = \lambda_{recon}\pounds_{recon} + \lambda_{cont} \pounds_{cont} + \lambda_{style} \pounds_{style}
\end{equation}
The loss hyper-parameters are set as follows: $\lambda_{recon}=1.0$, $\lambda_{cont}=0.5$, and $\lambda_{style}=0.5$.

\begin{table*}[htbp]
\renewcommand\arraystretch{1.5}

\centering
\caption{Quantitative results of the T2w images with different enlargement scales on the IXI-Test. The best and second-best are marked in red and blue, respectively.} 
\label{table:resultsT2} 
\begin{tabular}{r|m{0.8cm}m{0.8cm}m{0.8cm}m{0.8cm}m{0.8cm}m{0.8cm}|m{0.8cm}m{0.8cm}m{0.8cm}m{0.8cm}m{0.8cm}m{0.8cm}}
\hline
Dataset   & \multicolumn{12}{c}{IXI-Test}          \\
\hline
Scale     & \multicolumn{6}{c|}{$\times$4}&  \multicolumn{6}{c}{$\times$8} \\
\hline
Metrics   & PSNR  & SSIM   & Dice-W  & Dice-G   & Dice-CSF  & Dice-Avg    & PSNR  & SSIM & Dice-W  & Dice-G   & Dice-CSF  & Dice-Avg  \\
\hline
\hline
EDSR      & 32.94 & 0.9510 & 0.9716 & 0.9698   & 0.9377      & 0.9597        &   26.54    & 0.8566  & 0.9271 & 0.9170  & 0.9126      & 0.9198     \\
GFP-GAN   & 35.72     &  0.9761 & 0.9739   & 0.9654  &0.9488      & 0.9627        &   29.36    & 0.9259 &\textcolor{blue}{0.9440}  & 0.9103   & 0.9124      & 0.9223         \\
MINet    &   37.16     &  0.9812   & 0.9749 & \textcolor{blue}{0.9790}  & \textcolor{red}{0.9697}      & 0.9745        &   30.26    & 0.9430  & 0.9402 & \textcolor{blue}{0.9407}  & \textcolor{red}{0.9455}      & \textcolor{red}{0.9421}            \\
SwinIR    & 36.73     &  0.9815 & \textcolor{red}{0.9782} & 0.9714  & 0.9676      & 0.9724        &   29.95    & 0.9395   & 0.9382 & 0.9324 & \textcolor{blue}{0.9435}     & 0.9380        \\
Restormer    & \textcolor{blue}{37.25}     &  \textcolor{blue}{0.9836}  & 0.9751 & \textcolor{red}{0.9794}  & 0.9683      & 0.9743        &   \textcolor{blue}{30.38}    & \textcolor{blue}{0.9440}  & 0.9405 & \textcolor{red}{0.9397}  & 0.9415      & 0.9405        \\
\hline
\textbf{TransMRSR}   & \textcolor{red}{37.37} &  \textcolor{red}{0.9843}  & \textcolor{blue}{0.9769}    & \textcolor{red}{0.9794}  & \textcolor{blue}{0.9688} &\textcolor{red}{0.9750}&   \textcolor{red}{30.42}   & \textcolor{red}{0.9443} & \textcolor{red}{0.9477}  & 0.9369  & 0.9398      & \textcolor{blue}{0.9415}        \\
\hline
\end{tabular}
\end{table*}

\section{Experiments}
\subsection{Datasets}
The training dataset used in this paper is from the IXI dataset. The IXI dataset consists of multiple modalities acquired from 576 subjects with $1\times 1\times 1$mm resolution. We perform the generation task on the multi-modal dataset containing T1-weighted images (T1w) and T2-weighted (T2w) images. We train SR models on the T1 images and T2 images respectively. The 576 volumes are split into two groups, 536 for pretraining StyleSwin and 40 for the super-resolution task. The 40 volumes are further divided into 25, 5, and 10 as training, validation, and test data respectively. Due to the 2D nature of the proposed method, we get 3696, 744, and 1481 images for training, validation, and testing correspondingly. Additionally, we randomly select 10 volumes from an in-house 3D 
T1w dataset named SixP, and an in-house 2D T1 dataset named Snata for testing. Image registration is performed with FSL in the MNI space. The size of each 3D volume is reoriented to $182\times218\times181$. we restore image x-z plane slices and discard slices without any information.

Our TransMRSR is trained on synthetic data. We downsample the HR images by factors of $r={4,8}$ in the $z$-axis for T1w images and in the $x$-axis for T2w images to simulate thick-slice MRI following \cite{zhao2020smore}. The thick-slice MRI then is upsampled to the original resolution for HR space restoration. We extract LR-HR training pairs along $x$-axis, \textit{i.e.}, x-z plane slices, for t1w volumes and $z$ for t2w volumes, \textit{i.e.}, x-y plane slices. The same downgrading operation is also used on the SixP dataset.

\subsection{Evaluation Metrics}
First, we assess the effectiveness of models utilizing indices of Peak Signal-to-Noise Ratio (PSNR) and Structural Similarity (SSIM). Second, we exmaine the SR results of the proposed TransMRSR and SOTA methods in downstream segmentation tasks. The spatial similarity of the reconstructed brain between reconstructed images and ground truth is measured using the Dice coefficient:
\begin{equation}
    Dice(X,Y)=\frac{2\left | X\cap Y  \right | }{\left | X \right |+\left | Y \right |  } 
\end{equation}
Specifically, we compute dice coefficient scores
on the grey matter (Dice-G), white
matter (Dice-W), cerebrospinal fluid (Dice-CSF), and average dice score (Dice-Avg).

\begin{figure*}
\begin{center}
    \includegraphics[width=1.9\columnwidth]{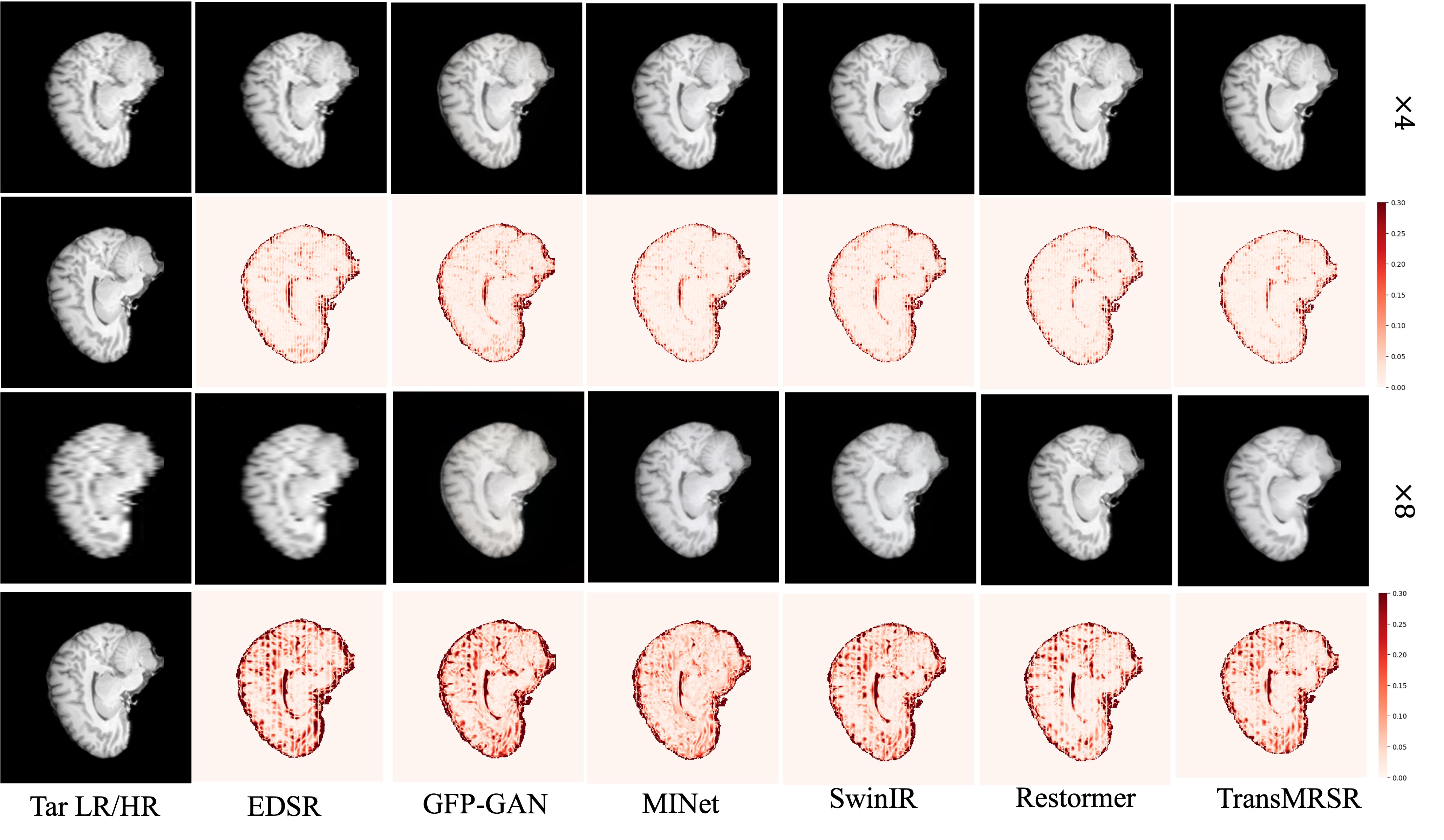}
\end{center}
\caption{Visual comparison of different SR reconstruction methods on the IXI test dataset with $\times 4$ and  $\times 8$ enlargement. The reconstructed images and the corresponding error map are provided.}
\label{fig:ixi}

\end{figure*}

\begin{figure*}
\begin{center}
    \includegraphics[width=1.9\columnwidth]{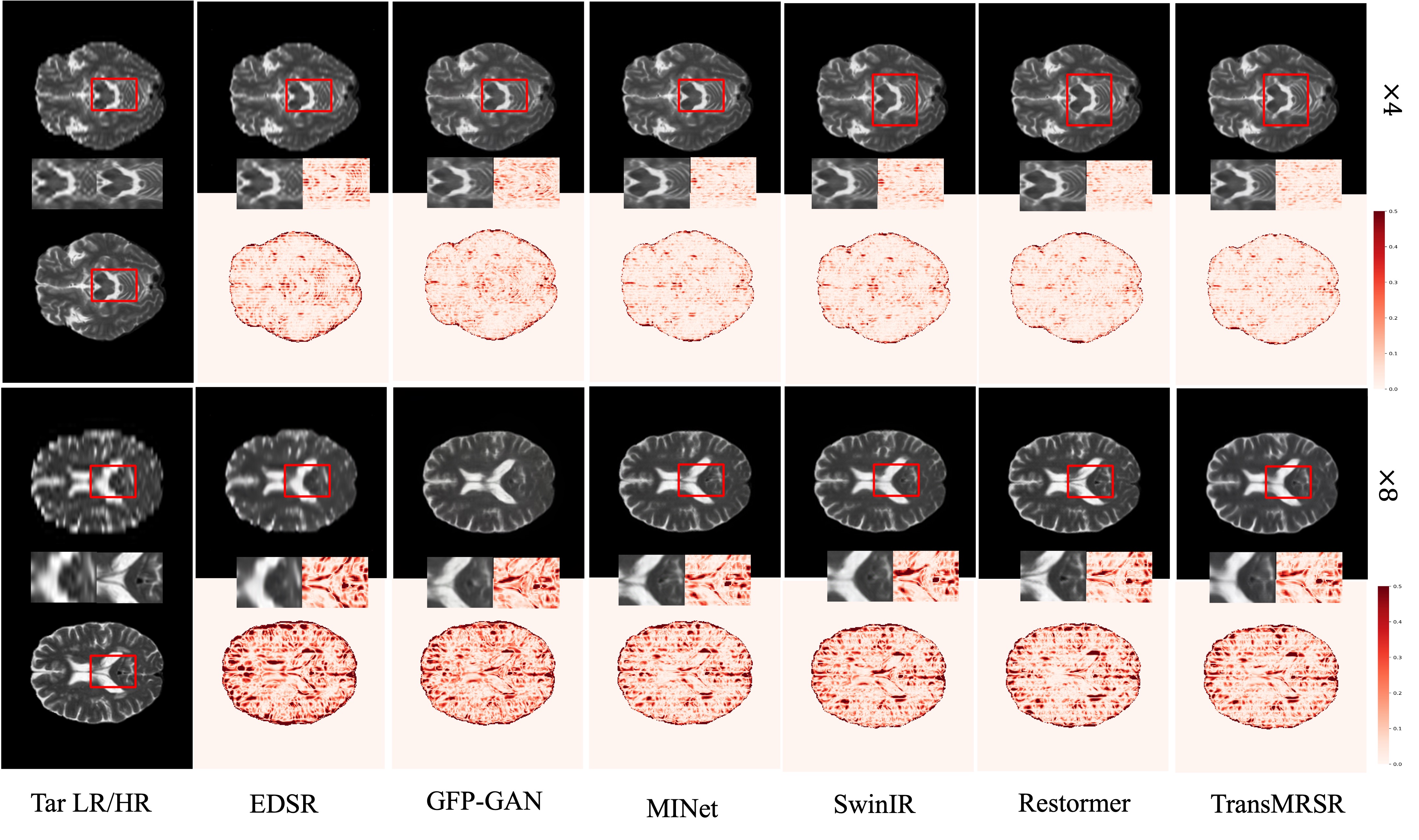}
\end{center}
\caption{Visual comparison of different SR reconstruction methods on the IXI test dataset with $\times 4$ and  $\times 8$ enlargement. The reconstructed images and the corresponding error map are provided.}
\label{fig:ixi_t2}

\end{figure*}
\begin{figure*}
\begin{center}
    \includegraphics[width=1.9\columnwidth]{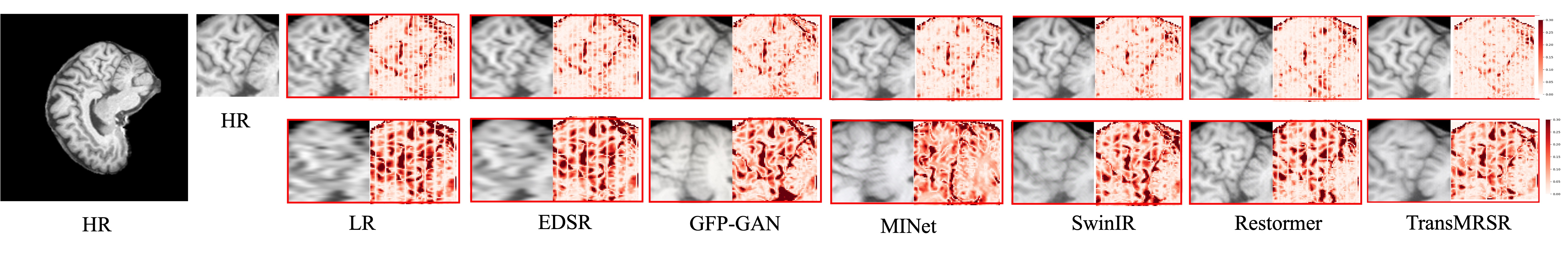}
\end{center}
\caption{Qualitative results of different SR reconstruction methods on the SixP dataset with $\times 8$ enlargement. The reconstructed images are zoomed in with different maps on the right for best view.}
\label{fig:liuyuan}
\label{Quali comparis}

\end{figure*}

\begin{figure*}
\begin{center}
    \includegraphics[width=1.9\columnwidth]{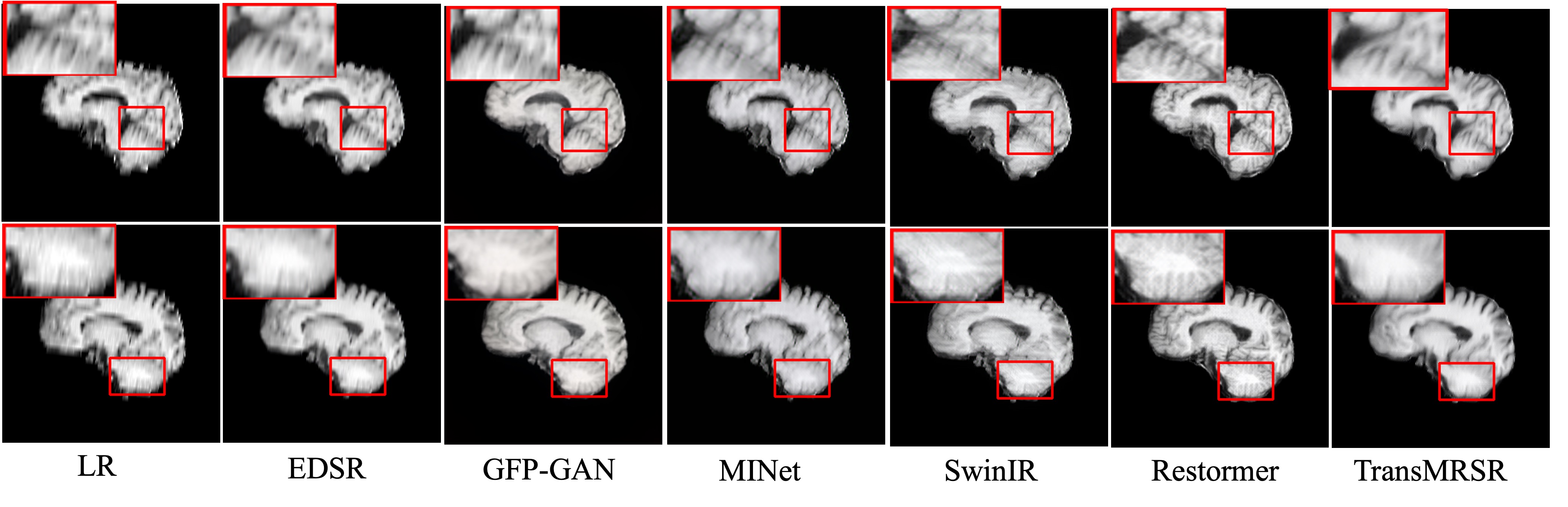}
\end{center}
\caption{Comparisons on the real-world SnaTa dataset. Images are best viewed when magnified.}
\label{fig:hualiao}
\end{figure*}
\subsection{Implementation Details}
We first extract HR slices from x-y, x-z, and y-z planes and pad them to $256 \times 256$ for pre-trained StyleSwin. The channel multiplier of StyleSwin is set to one for the compact model. We use two residual blocks to extract shallow features. The UNet consists of six upsamples and six downsamples, each with two successive swin transformer block \cite{liu2021swin}. The KMeans algorithm is employed to obtain 8 cluster centers of 60,000 randomly sampled latent codes. We find the nearest center for each latent code using euclidean distance. 

We train our model using two NVIDIA Tesla V100 GPUs based on PyTorch framework. We pad the image size to $256\times256$ as input. We augment the training data with random horizontal flip, vertical flip, and transpose in H and W dimensions. The training batch size is set to 6. The learning rate starts from 0.1 and decreases by a factor of 0.5 every 30 epochs. We use  Adam optimizer with $\beta_{1}=0.5$, $\beta_{2}=0.999$. The maximum number of training rounds is 100. When the PSNR obtained in the validation set is continuously increased by no more than 0.05 for 10 rounds, we terminate it.

\subsection{Main Results}
We compare our model with five SR methods: EDSR \cite{lim2017enhanced}, GFP-GAN \cite{wang2021towards}, MINet \cite{feng2021multi}, SwinIR \cite{liang2021swinir} and Rest-ormer \cite{zamir2022restormer} through both quantitative and qualitative results. For a fair comparison, we remove the last upsampling module of all networks. Images are upsampled to high-resolution space before passing through the network.


\subsubsection{Quantitative Results}
We compute PSNR/SSIM scores with three 
T1w datas-ets under $4\times$ and $8\times$ enlargement. As shown in Table. \ref{table:results}, our model achieves the best performance on all datasets. We further demonstrate the power of rich prior to preserve texture and structure information under $8 \times$ enlargement. Our model outperforms the best model by 0.3 dB and 0.0012 in PSNR and SSIM respectively. To assess the reliability of the reconstructed images, we perform the segmentation task with a Brain MRI analysis tool named FSL. We excluded segmentation results on images reconstructed by EDSR due to the poor quality. As shown in Fig. \ref{fig:dice}, Our TransMRSR produces consistent improvement in the downstream task. The Dice coefficient scores of the Gray matter, the Cerebrospinal fluid, and the average dice scores are the highest. However, FSL takes almost all voxels as white matter resulting in higher Dice-W, lower Dice-G, Dice-CSF, and Dice-Avg values for mediocre models (\textit{i.e}, GFP-GAN).

In addition, we apply the first-stage multimodal generative priors to perform super-resolution tasks on the IXI T2w dataset. Performance measurements are listed in Table. \ref{table:resultsT2}.  The superior performance demonstrates that the generative priors have the potential to handle multimodal SR tasks. As can be seen in Table \ref{table:resultsT2}, TransMRSR achieves consistent state-of-the-art performance on PSNR and SSIM metrics.

\subsubsection{Qualitative results}
Fig. \ref{fig:ixi} and Fig. \ref{fig:ixi_t2} provide the $\times 4$ and $\times 8$ enlargement of T1w and T2w images on the IXI test dataset. The darker colors in the error maps represent the larger errors. As can be seen, our TransMRSR restores most of the brain structures for multimodal images. The distribution of the brain sulci and gyri in the reconstructed images is basically consistent with the HR images. 

To evaluate the generalization capability and robustness, we further feed a synthetic and a real-world dataset into models.  As shown in Fig. \ref{fig:liuyuan}, our method can restore more anatomical details in the SixP dataset even on $\times 8$ SR tasks. The model used to restore real-world images is trained on the IXI dataset under $\times 8$ enlargement. Fig. \ref{fig:hualiao} shows TransMRSR restores more details with less noise and artifacts compared with other networks. Despite being trained on a synthetic dataset, our TGPSRMR can handle complex degradations.

\subsection{Ablation study}
In this section, we conduct an ablation study on the $8\times$ SR task. We construct three variant models: w/o GP, which is our model without generative prior, w/o SDT, which is our model without self-distilled truncation trick, w/o MRSE, which is our model without multi-resolution encoder features (i.e, $F_{l}$) fed into the decoder, and w/o skip connection, which is our model without skip connection between shallow feature and deep feature, the input and the output of the deep feature extraction module. Testing is performed on the IXI test dataset. Next, we demonstrate the effectiveness of key components of TransMRSR.
\subsubsection{Improvements of Generative Prior (GP)}
\begin{figure*}
\begin{center}
    \includegraphics[width=1.9\columnwidth]{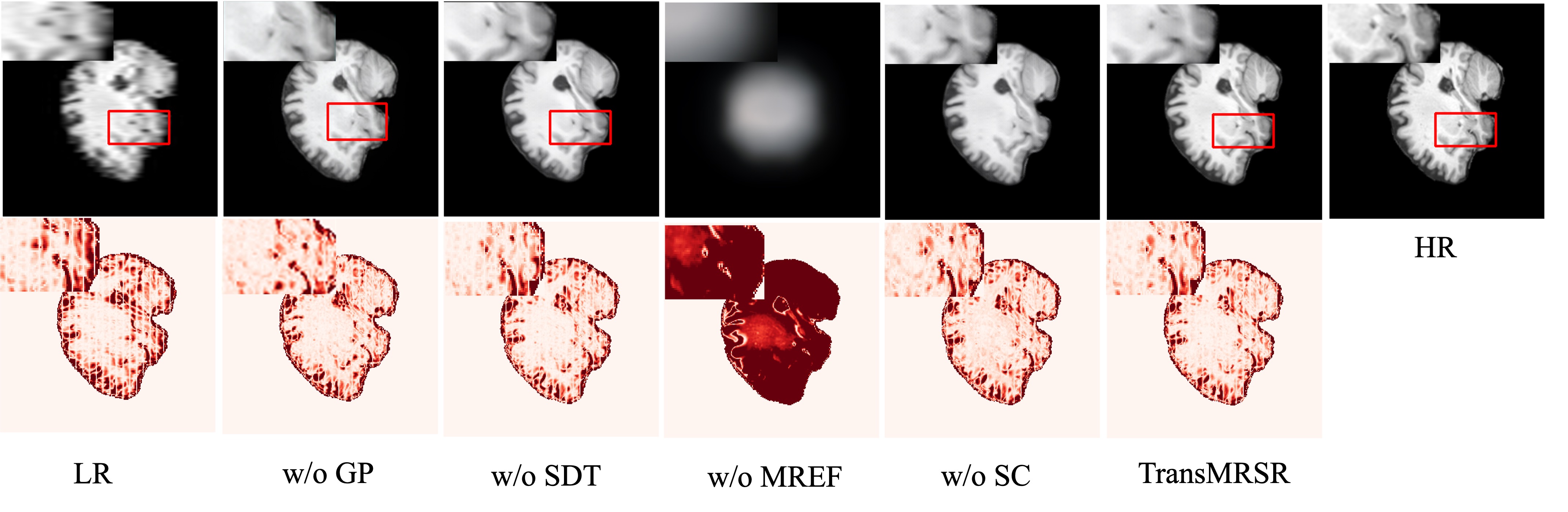}
\end{center}
\caption{Qualitative comparison on the different variant model under IXI-Test dataset with $\times 8$ enlargement scale. The first row includes the reconstructed images and the second row is the corresponding error maps. Zoom in for best view.}
\label{fig:ablation}
\end{figure*}

As shown in Table. \ref{Table:ablation}, our TransMRSR obtains a 0.63dB improvement in PSNR and a 0.0067 boost in SSIM compared to a non-pretrained decoder. The structural prior of the brain is preserved in pre-trained StyleSwin so that the network can retain more structural information under large-scale SR tasks, as shown in Fig. \ref{fig:ablation}. To demonstrate the effectiveness of fine-tuning the generative prior, we compare the performance of freezing versus fine-tuning the parameters of StyleSwin. As shown in Table. \ref{Table:fine-tuned}, Our fine-tuning strategy performs better (0.31dB+ PSNR, 0.0041+ SSIM) when restoring T2 images.

\begin{table*}[]
\renewcommand\arraystretch{1.5}
\centering
\caption{Ablation study results on IXI-Test under $\times 8$ enlargement scale. The best is marked in red.} 
\label{Table:ablation} 
\begin{tabular}{c|c|c|c|c|c|c|c|c|c|c}
\hline
\multirow{2}{*}{Configuration} & \multicolumn{4}{c|}{Modules}  & \multicolumn{6}{c}{Metrics}\\
\cline{2-11}
              & GP   & SDT   & MREF & SC  & PSNR    & SSIM  & Dice-W &Dice-G & Dice-CSF & Dice-Avg\\
\hline
\hline
w/o GP        & $\times$  & $\times$     & $\surd$   & $\surd$    & 30.33   & 0.9351 & \textcolor{red}{0.9521} & 0.9062 & 0.8690 &0.9091 \\
\hline
w/o SDT      & $\surd$    & $\times$    & $\surd$    & $\surd$    & 30.70&0.9431&0.9441&0.9414&0.9601&0.9485\\
\hline
w/o MREF      & $\surd$    & $\surd$    & $\times$    & $\surd$    &  14.34       &  0.4041&0.6962 & 0.6123 & 0.9118 & 0.7401     \\
\hline
w/o SC        & $\surd$     & $\surd$   & $\surd$    & $\times$  & 30.36   & 0.9140 & 0.9473 & 0.9427 & 0.9512 &0.9471\\
\hline
\textbf{TransMRSR}       & $\surd$    & $\surd$    & $\surd$     & $\surd$   & \textcolor{red}{30.85}   & \textcolor{red}{0.9451} & 0.9381 & \textcolor{red}{0.9479} & \textcolor{red}{0.9673} & \textcolor{red}{0.9511}\\
\hline
\end{tabular}
\end{table*}

\subsubsection{Effectiveness of Self-distilled Truncation (SDT)}
 A performance drop is observed if we do not use self-distilled truncation trick for both T1w and T2w restoration (see Table. \ref{table:resultsT2} and Table. \ref{Table:sdt})  The SDT trick transfers knowledge from multimodal generative tasks to multimodal super-resolution tasks when we perform only one generative task to encapsulate multimodal priors. Specifically, the help of the nearest latent centroid is obvious in the restoration of T2w images, resulting in an improvement of 0.17dB and 0.001 in PSNR and SSIM respectively.
\begin{table}[]
\renewcommand\arraystretch{1.5}
\centering
\caption{Ablation study results on training strategy about GP under $\times 8$ enlargement scale.} 
\label{Table:fine-tuned} 
\begin{tabular}{c|cc|cc}
\hline

\multirow{2}{*}{Configuration} &  \multicolumn{2}{c|}{IXI-Test(T1)}   & \multicolumn{2}{c}{IXI-Test(T2)}\\
\cline{2-5}
            &PSNR   & SSIM &PSNR   & SSIM \\
\hline
\hline
forzen GP        & 30.84  & 0.9450  & 30.11 & 0.9402\\
fine-tuned GP     & 30.85    &0.9451 & 30.42 & 0.9443 \\

\hline
\end{tabular}
\end{table}

\begin{table}[]
\renewcommand\arraystretch{1.5}
\centering
\caption{Ablation study results on SDT under $\times 8$ enlargement scale.} 
\label{Table:sdt} 
\begin{tabular}{c|cc|cc}
\hline

\multirow{2}{*}{Configuration} &  \multicolumn{2}{c|}{IXI-Test(T1)}   & \multicolumn{2}{c}{IXI-Test(T2)}\\
\cline{2-5}
            &PSNR   & SSIM &PSNR   & SSIM \\
\hline
\hline
w/o SDT        & 30.70  & 0.9431  & 30.25 & 0.9433\\
with SDT     & 30.85    &0.9451 & 30.42 & 0.9443 \\

\hline
\end{tabular}
\end{table}
\subsubsection{Impact of multi-resolution encoder features (MREF)}
When we remove all encoder features, the network resembles the typical GAN inversion methods that only learn latent code. Table. \ref{Table:ablation} shows the worst performance when we only preserve the latent codes to guide the decoder.
As shown in Fig. \ref{fig:ablation}, the restored image is of low quality and cannot show any brain structures. It is consistent with GFP-GAN \cite{wang2021towards}, the low-dimension latent vectors are insufficient to guide the restoration. We find a huge improvement (PSNR: 23.32dB, SSIM: 0.8187) when we fix generative priors without MREF. 
\subsubsection{Importance of skip connection (SC)}
we investigate Tran without skip connection between shallow features and deep features. The network directly learns the recovery of HR images instead of the residual image. As shown in Table. \ref{Table:ablation} and Fig. \ref{fig:ablation}, a performance drop is observed and boundaries between brain structures smooth out.

\section{Conclusion}
In this paper, we propose a transformer-based framework named TransMRSR for brain MRI SR reconstruction. Our model is able to restore fine details even on large factor reconstruction tasks, \textit{e.g.}, $\times 8$, with the help of generative prior encapsulated on a pre-trained StyleGAN. Extensive experiments show that TransMRSR outperforms other competing methods in terms of visual quality and quantitative results. Specifically, the degree of folding of the sulci and gyri in the image restored by our network is very close to that in the ground-truth. The robustness of our model to recover real-world dataset makes it potentially suitable for clinical applications. Future work includes measuring the authenticity of the restored images and investigating the effect of TransMRSR on downstream analysis tasks, such as lesion segmentation. Besides, we shall design a better way of knowledge distillation to exploit generative priors for more medical image tasks and methods.
\label{sec:5}
\\
\\
\textbf{Funding} This work was supported by Clinical Special Program of Shanghai Municipal Health Commission under Grant 2022404, and supported in part by the Shanghai Pujiang Program under Grant 22PJ1406800.
\\
\\
\textbf{Conflict of interest} The authors declare that they have no conflict of
interest.
\\
\\
\textbf{Data availability} The IXI dataset is publicly available at \href{https://brain-development.org/ixi-dataset}{https://brain-development.org/ixi-dataset}, and the private dataset SixP and Snata datasets are available from the corresponding author Bin Sheng on reasonable request.


\bibliographystyle{IEEEtran}
\bibliography{references}

\begin{thebibliography}{10}
\providecommand{\url}[1]{#1}
\csname url@samestyle\endcsname
\providecommand{\newblock}{\relax}
\providecommand{\bibinfo}[2]{#2}
\providecommand{\BIBentrySTDinterwordspacing}{\spaceskip=0pt\relax}
\providecommand{\BIBentryALTinterwordstretchfactor}{4}
\providecommand{\BIBentryALTinterwordspacing}{\spaceskip=\fontdimen2\font plus
\BIBentryALTinterwordstretchfactor\fontdimen3\font minus
  \fontdimen4\font\relax}
\providecommand{\BIBforeignlanguage}[2]{{%
\expandafter\ifx\csname l@#1\endcsname\relax
\typeout{** WARNING: IEEEtran.bst: No hyphenation pattern has been}%
\typeout{** loaded for the language `#1'. Using the pattern for}%
\typeout{** the default language instead.}%
\else
\language=\csname l@#1\endcsname
\fi
#2}}
\providecommand{\BIBdecl}{\relax}
\BIBdecl

\bibitem{xia2021super}
Y.~Xia, N.~Ravikumar, J.~P. Greenwood, S.~Neubauer, S.~E. Petersen, and A.~F.
  Frangi, ``Super-resolution of cardiac mr cine imaging using conditional gans
  and unsupervised transfer learning,'' \emph{Medical Image Analysis}, vol.~71,
  p. 102037, 2021.

\bibitem{zhao2020smore}
C.~Zhao, B.~E. Dewey, D.~L. Pham, P.~A. Calabresi, D.~S. Reich, and J.~L.
  Prince, ``Smore: a self-supervised anti-aliasing and super-resolution
  algorithm for mri using deep learning,'' \emph{IEEE transactions on medical
  imaging}, vol.~40, no.~3, pp. 805--817, 2020.

\bibitem{liu2021recycling}
G.~Liu, Z.~Cao, Q.~Xu, Q.~Zhang, F.~Yang, X.~Xie, J.~Hao, Y.~Shi, B.~C.
  Bernhardt, Y.~He \emph{et~al.}, ``Recycling diagnostic mri for empowering
  brain morphometric research--critical \& practical assessment on
  learning-based image super-resolution,'' \emph{Neuroimage}, vol. 245, p.
  118687, 2021.

\bibitem{peng2020saint}
C.~Peng, W.-A. Lin, H.~Liao, R.~Chellappa, and S.~K. Zhou, ``Saint: spatially
  aware interpolation network for medical slice synthesis,'' in
  \emph{Proceedings of the IEEE/CVF Conference on Computer Vision and Pattern
  Recognition}, 2020, pp. 7750--7759.

\bibitem{zhang2021mr}
Y.~Zhang, K.~Li, K.~Li, and Y.~Fu, ``Mr image super-resolution with squeeze and
  excitation reasoning attention network,'' in \emph{Proceedings of the
  IEEE/CVF Conference on Computer Vision and Pattern Recognition}, 2021, pp.
  13\,425--13\,434.

\bibitem{chen2018brain}
Y.~Chen, Y.~Xie, Z.~Zhou, F.~Shi, A.~G. Christodoulou, and D.~Li, ``Brain mri
  super resolution using 3d deep densely connected neural networks,'' in
  \emph{2018 IEEE 15th international symposium on biomedical imaging (ISBI
  2018)}.\hskip 1em plus 0.5em minus 0.4em\relax IEEE, 2018, pp. 739--742.

\bibitem{du2020brain}
J.~Du, L.~Wang, Y.~Liu, Z.~Zhou, Z.~He, and Y.~Jia, ``Brain mri
  super-resolution using 3d dilated convolutional encoder--decoder network,''
  \emph{IEEE Access}, vol.~8, pp. 18\,938--18\,950, 2020.

\bibitem{wang2020enhanced}
J.~Wang, Y.~Chen, Y.~Wu, J.~Shi, and J.~Gee, ``Enhanced generative adversarial
  network for 3d brain mri super-resolution,'' in \emph{Proceedings of the
  IEEE/CVF Winter Conference on Applications of Computer Vision}, 2020, pp.
  3627--3636.

\bibitem{chen2018efficient}
Y.~Chen, F.~Shi, A.~G. Christodoulou, Y.~Xie, Z.~Zhou, and D.~Li, ``Efficient
  and accurate mri super-resolution using a generative adversarial network and
  3d multi-level densely connected network,'' in \emph{Medical Image Computing
  and Computer Assisted Intervention--MICCAI 2018: 21st International
  Conference, Granada, Spain, September 16-20, 2018, Proceedings, Part
  I}.\hskip 1em plus 0.5em minus 0.4em\relax Springer, 2018, pp. 91--99.

\bibitem{li2022transformer}
G.~Li, J.~Lv, Y.~Tian, Q.~Dou, C.~Wang, C.~Xu, and J.~Qin,
  ``Transformer-empowered multi-scale contextual matching and aggregation for
  multi-contrast mri super-resolution,'' in \emph{Proceedings of the IEEE/CVF
  Conference on Computer Vision and Pattern Recognition}, 2022, pp.
  20\,636--20\,645.

\bibitem{zhang2022styleswin}
B.~Zhang, S.~Gu, B.~Zhang, J.~Bao, D.~Chen, F.~Wen, Y.~Wang, and B.~Guo,
  ``Styleswin: Transformer-based gan for high-resolution image generation,'' in
  \emph{Proceedings of the IEEE/CVF conference on computer vision and pattern
  recognition}, 2022, pp. 11\,304--11\,314.

\bibitem{dong2015image}
C.~Dong, C.~C. Loy, K.~He, and X.~Tang, ``Image super-resolution using deep
  convolutional networks,'' \emph{IEEE transactions on pattern analysis and
  machine intelligence}, vol.~38, no.~2, pp. 295--307, 2015.

\bibitem{liu2018robust}
X.~Liu, L.~Chen, W.~Wang, and J.~Zhao, ``Robust multi-frame super-resolution
  based on spatially weighted half-quadratic estimation and adaptive btv
  regularization,'' \emph{IEEE Transactions on Image Processing}, vol.~27,
  no.~10, pp. 4971--4986, 2018.

\bibitem{liu2020end}
X.~Liu, L.~Kong, Y.~Zhou, J.~Zhao, and J.~Chen, ``End-to-end trainable video
  super-resolution based on a new mechanism for implicit motion estimation and
  compensation,'' in \emph{Proceedings of the IEEE/CVF Winter Conference on
  Applications of Computer Vision}, 2020, pp. 2416--2425.

\bibitem{wang2021single}
W.~Wang, J.~Hu, X.~Liu, J.~Zhao, and J.~Chen, ``Single image super resolution
  based on multi-scale structure and non-local smoothing,'' \emph{EURASIP
  Journal on Image and Video Processing}, vol. 2021, no.~1, p.~16, 2021.

\bibitem{shi2021learning}
Z.~Shi, X.~Liu, C.~Li, L.~Dai, J.~Chen, T.~N. Davidson, and J.~Zhao, ``Learning
  for unconstrained space-time video super-resolution,'' \emph{IEEE
  Transactions on Broadcasting}, vol.~68, no.~2, pp. 345--358, 2021.

\bibitem{liu2021exploit}
X.~Liu, K.~Shi, Z.~Wang, and J.~Chen, ``Exploit camera raw data for video
  super-resolution via hidden markov model inference,'' \emph{IEEE Transactions
  on Image Processing}, vol.~30, pp. 2127--2140, 2021.

\bibitem{chu2022nafssr}
X.~Chu, L.~Chen, and W.~Yu, ``Nafssr: stereo image super-resolution using
  nafnet,'' in \emph{Proceedings of the IEEE/CVF Conference on Computer Vision
  and Pattern Recognition}, 2022, pp. 1239--1248.

\bibitem{kim2016accurate}
J.~Kim, J.~K. Lee, and K.~M. Lee, ``Accurate image super-resolution using very
  deep convolutional networks,'' in \emph{Proceedings of the IEEE conference on
  computer vision and pattern recognition}, 2016, pp. 1646--1654.

\bibitem{shi2016real}
W.~Shi, J.~Caballero, F.~Husz{\'a}r, J.~Totz, A.~P. Aitken, R.~Bishop,
  D.~Rueckert, and Z.~Wang, ``Real-time single image and video super-resolution
  using an efficient sub-pixel convolutional neural network,'' in
  \emph{Proceedings of the IEEE conference on computer vision and pattern
  recognition}, 2016, pp. 1874--1883.

\bibitem{ledig2017photo}
C.~Ledig, L.~Theis, F.~Husz{\'a}r, J.~Caballero, A.~Cunningham, A.~Acosta,
  A.~Aitken, A.~Tejani, J.~Totz, Z.~Wang \emph{et~al.}, ``Photo-realistic
  single image super-resolution using a generative adversarial network,'' in
  \emph{Proceedings of the IEEE conference on computer vision and pattern
  recognition}, 2017, pp. 4681--4690.

\bibitem{feng2021multi}
C.-M. Feng, H.~Fu, S.~Yuan, and Y.~Xu, ``Multi-contrast mri super-resolution
  via a multi-stage integration network,'' in \emph{Medical Image Computing and
  Computer Assisted Intervention--MICCAI 2021: 24th International Conference,
  Strasbourg, France, September 27--October 1, 2021, Proceedings, Part VI
  24}.\hskip 1em plus 0.5em minus 0.4em\relax Springer, 2021, pp. 140--149.

\bibitem{lyu2020mri}
Q.~Lyu, H.~Shan, and G.~Wang, ``Mri super-resolution with ensemble learning and
  complementary priors,'' \emph{IEEE Transactions on Computational Imaging},
  vol.~6, pp. 615--624, 2020.

\bibitem{zhang2019mri}
H.~Zhang, H.~Li, D.~Zhang, Y.~Zhang, X.~Wang, Y.~Xia, Y.~Shi, and W.~Wang,
  ``Mri super-resolution using 3d deeply residual and densely convolutional
  neural networks,'' \emph{IEEE Transactions on Medical Imaging}, vol.~38,
  no.~1, pp. 167--179, 2019.

\bibitem{chartier2018multi}
S.~Chartier, A.~M. Khairy, M.~Reisert, S.~Meriaux, J.~Montagnat, and
  H.~Liebgott, ``Multi-scale 3d generative adversarial networks for mr image
  synthesis,'' in \emph{International Conference on Medical Image Computing and
  Computer-Assisted Intervention}.\hskip 1em plus 0.5em minus 0.4em\relax
  Springer, 2018, pp. 198--206.

\bibitem{jin2022deep}
Z.~Jin, Y.~Li, W.~Chen, H.~Liu, Y.~Zhang, and Q.~Zhang, ``Deep learning-based
  3d mri super-resolution with multiple inference paths,'' \emph{IEEE Journal
  of Biomedical and Health Informatics}, 2022.

\bibitem{bau2020semantic}
D.~Bau, H.~Strobelt, W.~Peebles, J.~Wulff, B.~Zhou, J.-Y. Zhu, and A.~Torralba,
  ``Semantic photo manipulation with a generative image prior,'' \emph{arXiv
  preprint arXiv:2005.07727}, 2020.

\bibitem{goetschalckx2019ganalyze}
L.~Goetschalckx, A.~Andonian, A.~Oliva, and P.~Isola, ``Ganalyze: Toward visual
  definitions of cognitive image properties,'' in \emph{Proceedings of the
  ieee/cvf international conference on computer vision}, 2019, pp. 5744--5753.

\bibitem{xia2022gan}
W.~Xia, Y.~Zhang, Y.~Yang, J.-H. Xue, B.~Zhou, and M.-H. Yang, ``Gan inversion:
  A survey,'' \emph{IEEE Transactions on Pattern Analysis and Machine
  Intelligence}, 2022.

\bibitem{menon2020pulse}
S.~Menon, A.~Damian, S.~Hu, N.~Ravi, and C.~Rudin, ``Pulse: Self-supervised
  photo upsampling via latent space exploration of generative models,'' in
  \emph{Proceedings of the ieee/cvf conference on computer vision and pattern
  recognition}, 2020, pp. 2437--2445.

\bibitem{gu2020image}
J.~Gu, Y.~Shen, and B.~Zhou, ``Image processing using multi-code gan prior,''
  in \emph{Proceedings of the IEEE/CVF conference on computer vision and
  pattern recognition}, 2020, pp. 3012--3021.

\bibitem{zhu2020domain}
J.~Zhu, Y.~Shen, D.~Zhao, and B.~Zhou, ``In-domain gan inversion for real image
  editing,'' in \emph{Computer Vision--ECCV 2020: 16th European Conference,
  Glasgow, UK, August 23--28, 2020, Proceedings, Part XVII 16}.\hskip 1em plus
  0.5em minus 0.4em\relax Springer, 2020, pp. 592--608.

\bibitem{chan2021glean}
K.~C. Chan, X.~Wang, X.~Xu, J.~Gu, and C.~C. Loy, ``Glean: Generative latent
  bank for large-factor image super-resolution,'' in \emph{Proceedings of the
  IEEE/CVF conference on computer vision and pattern recognition}, 2021, pp.
  14\,245--14\,254.

\bibitem{wang2021towards}
X.~Wang, Y.~Li, H.~Zhang, and Y.~Shan, ``Towards real-world blind face
  restoration with generative facial prior,'' in \emph{Proceedings of the
  IEEE/CVF Conference on Computer Vision and Pattern Recognition}, 2021, pp.
  9168--9178.

\bibitem{korkmaz2022unsupervised}
Y.~Korkmaz, S.~U. Dar, M.~Yurt, M.~{\"O}zbey, and T.~Cukur, ``Unsupervised mri
  reconstruction via zero-shot learned adversarial transformers,'' \emph{IEEE
  Transactions on Medical Imaging}, vol.~41, no.~7, pp. 1747--1763, 2022.

\bibitem{karras2019style}
T.~Karras, S.~Laine, and T.~Aila, ``A style-based generator architecture for
  generative adversarial networks,'' in \emph{Proceedings of the IEEE/CVF
  conference on computer vision and pattern recognition}, 2019, pp. 4401--4410.

\bibitem{mokady2022self}
R.~Mokady, O.~Tov, M.~Yarom, O.~Lang, I.~Mosseri, T.~Dekel, D.~Cohen-Or, and
  M.~Irani, ``Self-distilled stylegan: Towards generation from internet
  photos,'' in \emph{ACM SIGGRAPH 2022 Conference Proceedings}, 2022, pp. 1--9.

\bibitem{liang2021swinir}
J.~Liang, J.~Cao, G.~Sun, K.~Zhang, L.~Van~Gool, and R.~Timofte, ``Swinir:
  Image restoration using swin transformer,'' in \emph{Proceedings of the
  IEEE/CVF international conference on computer vision}, 2021, pp. 1833--1844.

\bibitem{johnson2016perceptual}
J.~Johnson, A.~Alahi, and L.~Fei-Fei, ``Perceptual losses for real-time style
  transfer and super-resolution,'' in \emph{Computer Vision--ECCV 2016: 14th
  European Conference, Amsterdam, The Netherlands, October 11-14, 2016,
  Proceedings, Part II 14}.\hskip 1em plus 0.5em minus 0.4em\relax Springer,
  2016, pp. 694--711.

\bibitem{gatys2016image}
L.~A. Gatys, A.~S. Ecker, and M.~Bethge, ``Image style transfer using
  convolutional neural networks,'' in \emph{Proceedings of the IEEE conference
  on computer vision and pattern recognition}, 2016, pp. 2414--2423.

\bibitem{liu2021swin}
Z.~Liu, Y.~Lin, Y.~Cao, H.~Hu, Y.~Wei, Z.~Zhang, S.~Lin, and B.~Guo, ``Swin
  transformer: Hierarchical vision transformer using shifted windows,'' in
  \emph{Proceedings of the IEEE/CVF international conference on computer
  vision}, 2021, pp. 10\,012--10\,022.

\bibitem{lim2017enhanced}
B.~Lim, S.~Son, H.~Kim, S.~Nah, and K.~Mu~Lee, ``Enhanced deep residual
  networks for single image super-resolution,'' in \emph{Proceedings of the
  IEEE conference on computer vision and pattern recognition workshops}, 2017,
  pp. 136--144.

\bibitem{zamir2022restormer}
S.~W. Zamir, A.~Arora, S.~Khan, M.~Hayat, F.~S. Khan, and M.-H. Yang,
  ``Restormer: Efficient transformer for high-resolution image restoration,''
  in \emph{Proceedings of the IEEE/CVF Conference on Computer Vision and
  Pattern Recognition}, 2022, pp. 5728--5739.

\end{thebibliography}

\end{document}